# Observation of Spin-Transfer Switching In Deep Submicron-Sized and Low-Resistance Magnetic Tunnel Junctions

Yiming Huai[a], Frank Albert, Paul Nguyen, Mahendra Pakala and Thierry Valet
*Grandis Inc., R & D Department, 1266 Cadillac Court, Milpitas, California, 95035*

The spin-transfer effect has been studied in magnetic tunnel junctions (PtMn/CoFe/Ru/CoFe/$Al_2O_3$/CoFe/NiFe) with dimensions down to 0.1x0.2 $\mu m^2$ and resistance-area product RA in the range of 0.5-10 $\Omega\mu m^2$ ($\Delta R/R$=1-20%). Current-induced magnetization switching is observed with a critical current density of about $8\times10^6$ $A/cm^2$. The attribution of the switching to the spin-transfer effect is supported by a current-induced $\Delta R/R$ value identical to the one obtained from the R versus H measurements. Furthermore, the critical switching current density has clear dependence on the applied magnetic field, consistent with what has been observed previously in the case of spin-transfer induced switching in metallic multilayer systems.

Magnetization switching induced by spin-polarized current has stimulated considerable interest in recent years due to its rich fundamental physics and potential for new magnetoelectronic applications. Low switching current density and high read signal are required for the application of the spin-transfer switching to non-volatile magnetic random access memory (MRAM). Most of the work to date, however, has focused on magnetic metallic multilayers, which require large currents applied in the current-perpendicular-to-plane direction but yield small resistance (R) and nominal magnetoresistance ($\Delta R/R$).[1] On the other hand, magnetic tunnel junctions (MTJ) have both high R and $\Delta R/R$, resulting in high signal output. In order to utilize MTJs in spin transfer based MRAM, however, requires an understanding of the limits of both the spin transfer effect and the electron transport properties of tunnel barriers used in MTJs.

We report the observation of the spin-transfer effect in low-resistance MTJs ( RA=0.5-10 $\Omega\mu m^2$) with dimensions down to 0.1x 0.2 $\mu m^2$. These deep submicron-sized MTJs minimize the Oersted (vortex) field contribution due to large vertical current through the MTJ pillars. [2,3] MTJ films Ta20/NiFeCr35/PtMn140/ CoFe20/Ru8 /CoFe22/ $Al_2O_3$/ CoFe10/NiFe20/Ta50 (in Å) were deposited in a magnetron sputtering cluster system and annealed at 250-270 $^oC$ for 10 hours. A thin tunneling barrier was formed by two-step natural oxidation of the pre-deposited Al layer in a pure oxygen atmosphere.[4] The MTJ films were subsequently patterned into deep submicron ellipse-shaped pillars using DUV photolithography combined with resist trimming and ion milling. The pillar dimensions and microstructures have been characterized by high-resolution transmission electron microscope (TEM). The cross sectional TEM micrograph of an MTJ sample (0.12 x 0.23 $\mu m^2$ ellipse), taken along the long axis, shows a continuous well-defined alumina barrier layer (see Fig. 1). The edges of the nano-pillar are also well defined, smooth and steep. The lateral dimension is closed to 0.22 $\mu m$, considering the small overlayer of insulating material $Al_2O_3$ at the edge of the nanopillar.

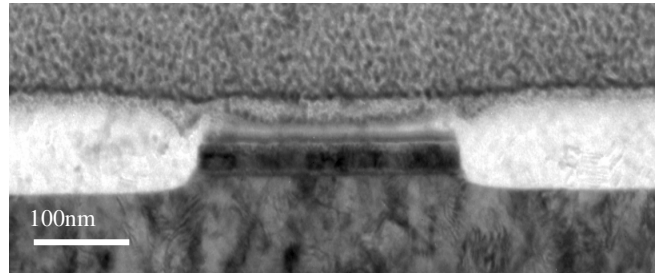

Fig. 1: Cross-sectional TEM micrograph of a sample with RA = 1.6 $\Omega\mu m^2$. The cross-section is taken along the long axis of the 0.12 x 0.23 $\mu m^2$ ellipse shaped nanopillar.

The resistance/magnetoresistance versus magnetic field and current were measured by a quasi-static tester with pulsed current capability. Breakdown voltages for the samples in the RA range studied here are found to be between 0.3-0.8 V, allowing a current flow of density up to $6\times10^7$ $A/cm^2$ without dielectric breakdown of the thin junction barriers.[5] The representative plots of resistance R (in the parallel state) versus the voltage bias, as shown in Fig. 2, exhibit two different types of behavior depending on RA value of the MTJ samples. For the low RA samples (0.2–1.6 $\Omega\mu m^2$), R increases with increasing voltage bias [see Fig. 2 (a)]. This R increase with voltage bias is similar to that characteristic of bottom spin-valve samples with similar structures except for the barrier layer, as shown in the insert of Fig. 2 (a) for comparison. For the higher RA samples (>1.6 $\Omega\mu m^2$), an inverse parabola similar to that seen previously in typical MTJs is observed [see Fig. 2 (b)],[3] suggesting that most of the current passes though the barrier by tunneling. The difference in the R versus voltage bias curves between the low RA and higher RA samples may be the result of incomplete oxidation of the barriers in the low RA samples, where the current passing across the barrier is mainly leakage current through the pinholes. For the low RA samples (<1.6 $\Omega\mu m^2$), the resistance becomes higher for higher voltage bias because of increasing electron-phonon and electron-magnon scatterings.

[a] Electronic mail: Yiming.Huai@GrandisInc.com

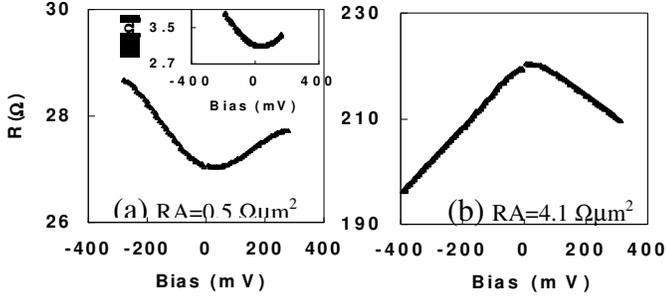

Fig. 2: Resistance versus voltage bias for (a) Low RA MTJ, (b) High RA MTJ. The insert in (a) shows R versus voltage bias for a bottom spin valve for comparison.

The resistance versus field scans are shown for two samples in Fig. 3 (a) and (c) having different RA, along with the corresponding resistance versus current scans in Fig. 3 (b) and (d). Positive field here is applied along the direction of the pinned sublayer (in the synthetic antiferromagnet) adjacent to the $Al_2O_3$ barrier layer. Positive direction of the pinned sublayer (in the synthetic antiferromagnet) adjacent to the $Al_2O_3$ barrier layer. Positive current I here denotes electron flow from the free to the pinned layer (current flow from the pinned to the free layer).

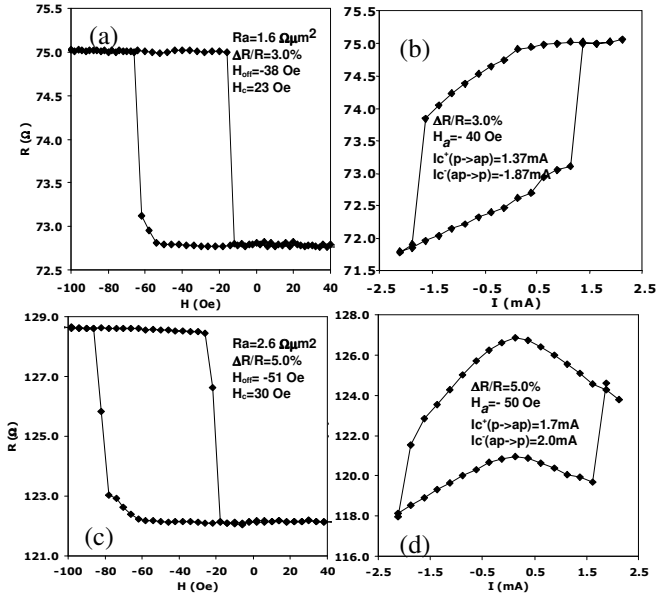

Fig. 3: DC resistance versus magnetic field scans [(a) and (c)] and corresponding resistance versus current scans [(b) and (d)] for two MTJ samples with RA=1.6 and 2.6 $\Omega\mu m^2$.

Two samples with RA of 1.6 and 2.6 $\Omega\mu m^2$ showed $\Delta R/R$= 3% and 5% (measured at low bias I=0.25 mA), respectively. In both samples, a ferromagnetic coupling offset field $H_{off}$, which arises from the orange-peel (Néel) coupling, can be seen. The R versus I scans were performed at zero effective magnetic bias with an applied field $H_a$ opposite and equal to the offset field $H_{off}$. The R versus I curves show sharp resistance transitions between parallel and anti-parallel magnetization alignments, exhibiting $\Delta R/R$ values identical to those obtained from R versus $H_a$ measurements. The average switching current density as calculated from ($I_c^+ - I_c^-$)/2A, where A is the junction area, $I_c^+$ and $I_c^-$ denotes the critical currents at which R jumps from low (parallel magnetizations) to high (antiparallel) and from high (antiparallel) to low (parallel), respectively, is around 8 x$10^6$ A/cm$^2$, comparable to those obtained from spin valves with same free layer structure. It should be pointed out that the current switching thresholds depend on the applied field, as shown in Fig. 4 for a MTJ sample with RA=2.6 $\Omega\mu m^2$. Because the torque due to the spin current must overcome the increasing torque due to the increased applied field, $I_c^-$ becomes more negative with a more negative $H_a$ (more negative $H_a$ favors more antiparallel magnetization alignments). The lack of $I_c^+$ data at less negative $H_a$ values is due to the limitation in the amount of current (-1.5-1.5 mA) that can be safely applied to MTJ sample during the R versus I scan without risking a dielectric breakdown of the junction barrier. Similar field dependence of the switching currents Ic has been observed in a number of MTJ samples. The insert in Fig. 4 shows Ic versus $H_a$ for a simple bottom spin valve with the same free layer structure for comparison. A kink is observed in the insert on the $I_c^-$ versus $H_a$ curve when $H_a$ approach the value that forces the free layer into alignment with the pinned layer in the absence of a current.

The clear field dependence of the critical switching current here is in contrast to the lack of field dependence of the current-induced switching observed in MTJs in earlier experiments due to the formation/annihilation of conduction channels by displacement of atoms or charges from the two electrodes into the thin insulating layer region (hot spots).[2,6] In these earlier experiments, the switching current was observed to be independent of applied field up to kOe range, well beyond the coercive field of the free layer.[2] And the $\Delta R/R$ observed during the current scans also varied over a wide range from below 50% up to 200% of the $\Delta R/R$ obtained from the R versus $H_a$ scan.[6] On the contrary, here we observe basically identical values of $\Delta R/R$ under both current and field scans, consistent with spin transfer experiments in Co/Cu/Co trilayers.[1] Furthermore, the critical switching current density in our MTJ samples is found to increase, while $\Delta R/R$ remains unchanged, with decreasing current pulse width (from 3000 ms to 3ms).[8] We found that $I_c$ decreases linearly with the logarithm of the pulse duration, a functional dependence which was previously observed in spin valve pillars,[9] and is expected for thermally activated switching.[9]

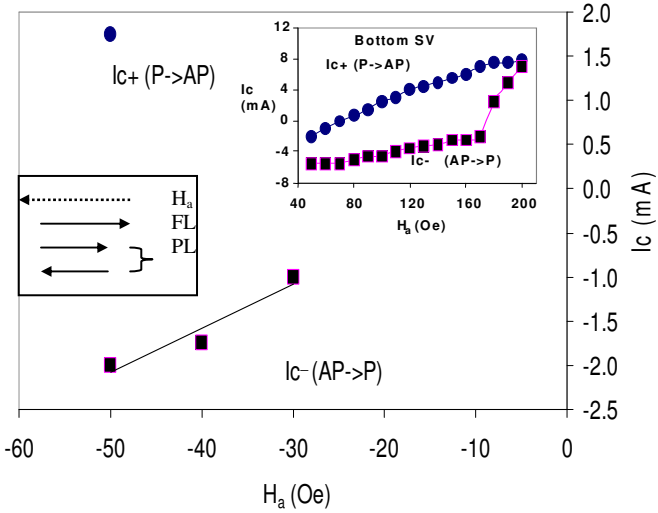

Fig. 4: Switching current versus external applied magnetic field $H_a$ for a MTJ with RA= 2.6 $\Omega\mu m^2$. The inert shows Ic versus $H_a$ for a simple bottom spin valve with the same free layer.

In addition to spin transfer switching, the R versus I curves shows distinct R versus voltage bias characteristics with a parabolic curve for RA<1.6 $\Omega\mu m^2$ and an inverse parabolic curve for RA >1.6 $\Omega\mu m^2$. It is interesting to note that the observed Ic values are similar for all the samples across the whole range of RA studied here (0.5-10 $\Omega\mu m^2$), even though a qualitative change in the electron transport process across the barrier layer is strongly indicated by the observed change in the curvature of the R versus voltage bias curves (see Fig.2). For the sample with RA=1.6 $\Omega\mu m^2$ shown in Fig.3b, the relatively flat curvature of the R versus I curve could be the result of the presence of both transport modes (electron tunneling across the barrier and leakage current through pinholes in the barrier layer).

We want to point out that the present experimental results open a new domain for the spin-transfer physics, and create new challenges in regard to providing an all encompassing theoretical framework. Transport in MTJ at finite bias involves a significant range of electronic state energies both above and below the Fermi level, as opposed to the situation in metallic systems where the transport is localized on the Fermi surface. Essentially all the theoretical models proposed so far to obtain the spin transfer torque from electronic transport calculations, either quantum mechanical [10,11,12] or semi-classical in nature, [14,15] have been derived in the limit of weak non-equilibrium. A proper generalization of these models in situations far from equilibrium is called for by the newly demonstrated spin-transfer effect in MTJ at finite current bias.

In conclusion, the spin-transfer effect has been observed in bottom synthetic MTJs with dimensions down to 0.1x 0.2 $\mu m^2$ and RA in the range of 0.5-10 $\Omega\mu m^2$ ($\Delta R/R$=1-20%). Spin transfer current induced switching was observed, as evidenced by R vs. H measurements compared with the current-induced $\Delta R/R$, along with the field dependence of the current driven switching. A qualitative difference in the electron transport behavior was observed in RA above and below 1.6 $\Omega\mu m^2$, however, spin transfer driven switching has been observed over a wide range of RA leading to the possibility of spin transfer based MRAM.

We would like to thank Xiaodong Dong, Yingjian Chen, Ki-Seok Moon, Changhe Chang, Wei Xiong, Zhitao Diao, Min Zhou and Henry Yuan for their helps in microfabrication and thin film processes.